\documentclass[%
  reprint,
  superscriptaddress,
  groupedaddress,
  amsmath,
  amssymb,
  aps,
  prab,
]{revtex4-2}

\usepackage{graphicx, svg, lipsum, verbatim, multirow, upgreek}
\usepackage{float, fdsymbol, amsthm, amssymb, algorithm, algpseudocode, amsmath, ragged2e}
\usepackage{wrapfig, placeins, subfigure}
\usepackage{dcolumn}
\usepackage{bm}
\usepackage{xcolor}
\usepackage{soul}
\usepackage{color}
\usepackage[normalem]{ulem}
\definecolor{mygreen}{rgb}{0,0.6,0}
\definecolor{myblue}{rgb}{0.1,0,0.8}
\definecolor{myred}{rgb}{0.9,0.2,0.2}

\usepackage[colorlinks=true,bookmarks=true,pdfborder={0 0 0},
    linkcolor=myblue,urlcolor=myblue,citecolor=myblue,
    breaklinks=true,bookmarksnumbered=true]{hyperref}

\begin{document}

\preprint{APS/123-QED}
\title{AI-Assisted Transport of Radioactive Ion Beams}

\author{S. Lopez-Caceres}
\author{D. Santiago-Gonzalez}
\affiliation{Physics Division, Argonne National Laboratory, Lemont, Illinois, USA}

\begin{abstract}
Beams of radioactive heavy ions allow researchers to study rare and unstable atomic nuclei, shedding light into the internal structure of exotic nuclei and on how chemical elements are formed in stars. 
However, the extraction and transport of radioactive beams rely on time-consuming expert-driven tuning methods, where hundreds of parameters are manually optimized. 
Here, we introduce a system that employs Artificial Intelligence (AI), specifically utilizing Bayesian Optimization, to assist in the transport process of radioactive beams. We apply our methodology to real-life scenarios showing advantages when compared with standard tuning methods. 
This AI-assisted approach can be extended to other radioactive beam facilities around the world to improve operational efficiency and enhance scientific output.

\end{abstract}
\maketitle
\section{\label{sec:intro}Introduction}
{
 \setlength{\parskip}{0pt}
  \setlength{\parindent}{10pt}

Particle accelerators continue to lead the way in scientific discovery and technological innovation, deepening our understanding of the universe, and driving advances that benefit society. 
By providing beams of particles, these accelerator facilities enable scientists to explore the fundamental building blocks of matter and unravel the mysteries of the cosmos. 
However, constructing and operating these accelerators is a complex undertaking, making them some of the most sophisticated scientific systems in existence.

Among these, facilities that produce radioactive ion beams (RIBs), see for example Refs.~\cite{Kub1992,Lie2001,savard_radioactive_2008,angert_gsi_2012,Moo2013,Wie2013,wrede_facility_2015,Bec2016,fraser_status_2017}, play a pivotal role. 
They allow researchers to study rare and unstable atomic nuclei, shedding light into the internal structure of exotic nuclei and on how elements are formed in stars~\cite{ye_physics_2025}. However, the standard processes for RIB production and transport rely on expert-driven manual methods that are both time-consuming and labor intensive, which can limit research productivity. 

In recent years, there has been increasing interest in using Artificial Intelligence (AI) applications and Machine Learning (ML) methods to improve the automation of accelerator operations. 
A review paper by Roussel et al.~\cite{roussel_bayesian_2024} highlights how Bayesian Optimization algorithms offer an effective and adaptable approach to addressing a wide range of optimization challenges in accelerator physics.

In this work, we introduce a system that uses AI to assist in the transport of radioactive ion beams. We present results from applying our methodology to real-life scenarios where the observable quantity to maximize is the activity from beta decay of stopped radioactive ions and where electrostatic devices are used to transport the beam. We also discuss the results and advantages when compared with standard tuning methods, and suggest that our method could be applied to similar radioactive beam facilities worldwide.

\section{\label{sec:bkg}Background and Motivation}
Since our method for AI-assisted radioactive beam transport was tested and implemented at the Californium Rare Isotope Breeder Upgrade (CARIBU)~\cite{savard_radioactive_2008} at Argonne National Laboratory, in the following paragraphs we occasionally refer to elements in that RIB facility. When possible, we discuss our method in general terms.  

At Argonne National Laboratory, the radioactive beam source plays a crucial role in advancing nuclear research as part of the Argonne Tandem-Linac Accelerator System (ATLAS). 
Since 2008, this source has been producing beams of radioactive heavy ions through the spontaneous fission of Californium-252 ($^{252}$Cf). 
These beams enable scientists to conduct groundbreaking experiments in nuclear structure, nuclear astrophysics, and national security.

To support this vital scientific work, the facility provides nearly 2,500 hours (approximately 100 days) of beam time per year to researchers from around the world. 
Each segment of the beamline comprises 5 to 15 variables that must be precisely coordinated to guide ions from the source to the experimental target station. 
With more than 20 beamline segments, this configuration results in a parameter space involving roughly 200 variables.

Currently, only experienced scientists, who have devoted considerable time to mastering these adjustments, perform the tuning to achieve optimal transmission efficiencies. 
While the scientific output has been significant, this manual approach poses limitations. Automating the tuning process could greatly enhance the facility's operational efficiency and scientific productivity by reducing the time and effort required for beam adjustments.
Moreover, because this tuning must be repeated whenever the beam is lost or switched to a different ion species, the demand on limited staff becomes even more pressing. 
Implementing automation would not only alleviate these challenges but also provide the flexibility needed to support a broader range of experiments and advancements in nuclear science.

\section{\label{sec:methodology}Methodology}

Given these technical challenges, the use AI applications, and in particular ML methods~\cite{edelen_opportunities_2018, scheinker_advanced_2020}, hold the potential to achieve complete automation of beam extraction, delivery, and optimization of related instruments. The aim of the work presented here is to automate radioactive beam tuning using ML techniques. Specifically, the use of Bayesian Optimization (BO) for control and tuning of beam elements \cite{roussel_bayesian_2024}. Although the underlying theory of this optimization approach is not new \cite{mockus_bayesian_1975}, computational advances and open-source software libraries \cite{pytorch_NEURIPS2019_9015, balandat2020botorch, tensorflow2015-whitepaper} have only recently matured enough to consider practical applications.

Bayesian Optimization is most effective in problems where the relation between variables is unknown or expensive to evaluate. 
A proxy model of probabilistic nature, Gaussian Process (GP) \cite{rasmussen_gaussian_2005}, acts as a surrogate to describe the functional relationship between parameters and observables. 
In combination with an acquisition function to find the next point to evaluate, this process is repeated until convergence is achieved. 
Adjusting the balance between exploration and exploitation can effectively maximize the optimal parameter space while searching for better results in new regions. 
With an augmented data set after each iteration, the result is a global optimum with the least number of iterative processes. 

Early successes in applying ML in the form of BO~\cite{shalloo_automation_2020, li_bayesian_2019, duris_bayesian_2020, mcintire_bayesian_2016, miskovich_online_2022, roussel_proximal_nodate, hao_reconstruction_2019} have proven the effectiveness of this method. 
This technique has been chosen over others such as Reinforcement Learning (RL) due to its non-parametric nature and minimal prior data requirement~\cite{wang_accelerator_2021, kaiser_reinforcement_2024}. GP does not require pre-training a Neural Network (NN) on either simulated or historical data. The learning is in real time from current samples, and the model estimations improve with each iteration. Although digital-twins of the system could be useful as a testbed for new algorithms, they fail to account for the underlying changing dynamics and fringe behaviors. Likewise, historical data is unable to accurately represent current parameters. Thus, BO is the most straightforward method to automate a complex, dynamic system -- like the transport of radioactive beams -- without \textit{a priori} knowledge.

\section{\label{sec:implementation}Implementation}
At RIB facilities, the successful transport of radioactive beams depends on the synchronized performance of numerous devices operating in optimum conditions to ensure maximum transport efficiency.
In Fig.~\ref{fig:nucaribu}, we show a diagram of the key elements involved in the transport of radioactive beams produced at CARIBU. 
After extraction from the ion source (bottom of the figure), electrostatic quadrupoles and steerers transport and focus ions into a set of dipole magnets and slits (labeled ``Isobar separator'') that act as a filter, only letting through ions with a particular mass-to-charge ratio. 
Quadrupoles focus ions vertically (Y) or horizontally (X), while steerers adjust beam direction in these planes. The beamline is divided into sections to facilitate the application of the method presented here.
Fission produces multiple species with different mass numbers (A), but experiments typically target isotopes with a specific A. 
Transporting all species can create background noise, complicating the ion identification. To filter unwanted fission products, the isobar separator is positioned early in the beamline for initial separation. 
Due to hysteresis in the magnets generating magnetic fields, this device is set to a specific field and does not require live tuning like other beam elements.

\begin{figure}[t]
    \centering       \includegraphics[width=1\linewidth]{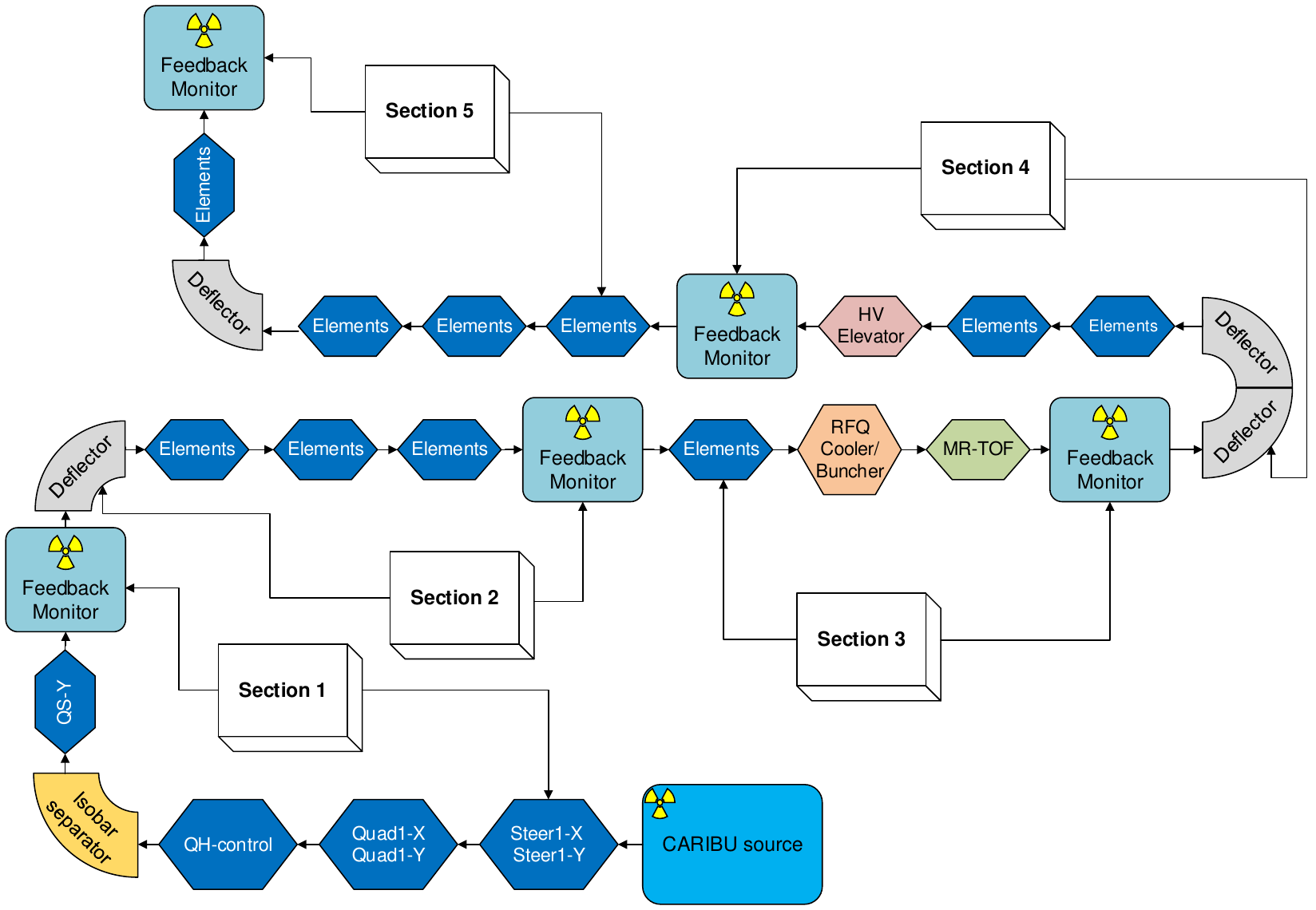}
    \caption{\justifying Key beam elements at CARIBU. Radioactive ions are produced at the source and transported through various elements, with arrows indicating the transport direction. Feedback monitors along the beamline measure rates, and BO is applied sequentially to each section.}
    \label{fig:nucaribu}
\end{figure}

Given the potential effectiveness of BO for these systems, the recently launched Badger optimizer~\cite{Badger} was selected for this work. This framework combines modern algorithms from the Xopt package~\cite{Xopt} with a graphical user interface (GUI) for customizing run parameters and tracking optimization progress. Additionally, a plugin mode facilitates communication with devices and instruments, allowing exploration of tailored operation sequences. A centralized directory for archiving optimization routines and past results enhances the platform's suitability for daily operations.

To integrate this optimization framework into the facility, beam elements and feedback monitors must be remotely accessible via a web server, a connection typically managed by the Accelerator Controls Group for security. However, due to safety concerns, the server must be enabled and disabled by an accelerator operator to act as the human-in-the-loop. Figure~\ref{fig:workflow} describes the basic workflow of our approach. Once the server is enabled and a secure connection is established, we use the developed functions to read (get) and write (set) values from and to the devices. 

\begin{figure}[t]
    \centering    \includegraphics[width=0.6\linewidth]{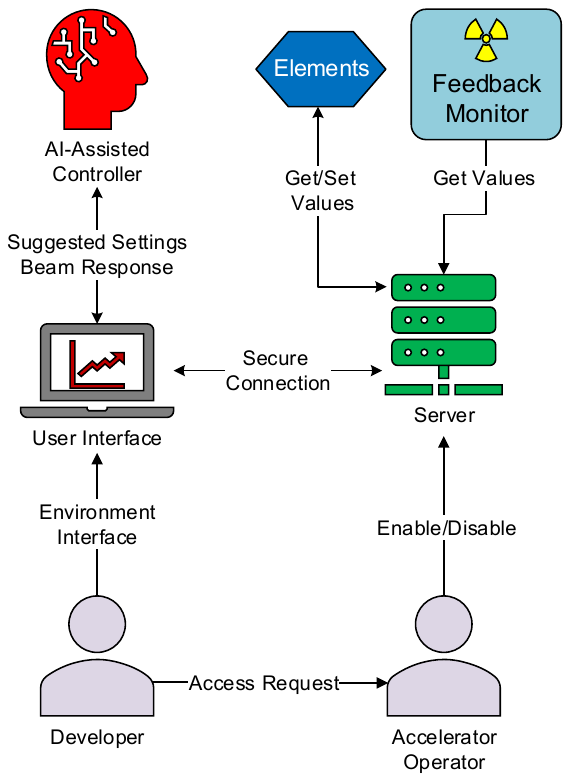}
    \caption{\justifying Workflow design for AI-assisted radioactive beam transport. Developers write the \textit{environment} and \textit{interface} plugins to communicate with the machines. Inside the accelerator network, operators enable a server to facilitate remote control of beamline elements.}
    \label{fig:workflow}
\end{figure}

As described in Ref.~\cite{Badger}, variables, variable ranges, and the observables are specified in an abstract element called \textit{Environment}. Variables, such as DC voltages controlling steerers, quadrupoles, and focusing elements, are set within operational limits. Historical ``good-tune'' values are loaded at initialization. The observable, defined as the optimization target (labeled ``Feedback'' in Fig.~\ref{fig:workflow}), uses signals from particle detectors. In the case of RIBs produced by CARIBU, we used Silicon detectors to observe electrons from the beta decay of the isotope of interest. Three high-level class methods facilitate communication with devices: reading current values, setting new values from optimization, and obtaining feedback for optimization. A second abstract element, called \textit{Interface}, contains the instructions for device communication.  

This approach offers extensive flexibility in designing procedural steps for machine communication. In case of a beam-down event, the optimizer can be re-started using the most recent optimal parameters.
This approach uses parameters that reflect the current beamline dynamics, rather than outdated historical data. The same interface can be applied to different beamline sections without altering the optimization source code.

Based on the known physical behavior of the beamline elements, the GP model employed a scaled Mat\'ern 5/2 kernel \cite{matern_spatial_1986, stein_interpolation_1999, rasmussen_gaussian_2005}. It incorporates BoTorch's default Gamma priors for the lengthscale $\ell$ and for the output scale $\sigma$ \cite{balandat2020botorch}, along with Automatic Relevance Determination (ARD) to learn a separate lengthscale $\ell_i$ for each input variable $x_i$ in the input vectors $\mathbf{x}, \mathbf{x'} \in \mathbb{R}^d$:
\begin{gather}   
k_{\text{Scaled}}(\mathbf{x}, \mathbf{x'}) = \sigma^2 \cdot k_{\text{Mat\'ern5/2}}(\mathbf{x}, \mathbf{x'}) 
\end{gather}
\begin{gather}
k_{\text{Mat\'ern5/2}}(\mathbf{x}, \mathbf{x'}) = \left(1 + \sqrt{5} \, r + \frac{5}{3} r^2 \right) \exp(-\sqrt{5} \, r) 
\end{gather}
\begin{gather}   
r = \sqrt{ \sum_{i=1}^d \left( \frac{x_i - x_i'}{\ell_i} \right)^2 } 
\end{gather}
\begin{gather}    
\ell_i \sim \text{Gamma}(3.0, 6.0), \quad \sigma^2 \sim \text{Gamma}(2.0, 0.15)
\end{gather}

This kernel was preferred over the Radial Basis Function (RBF) for its flexibility in modeling smooth functions. The Upper Confidence Bound (UCB) was chosen as the acquisition function due to its effective balance between exploration and exploitation, which is critical for optimizing complex systems. This function is defined as, 
\begin{gather}
\alpha(x) = \mu(x) + \beta \nu(x)
\label{eq:acq funct}
\end{gather}
\noindent where $\mu(x)$ is the predicted mean and $\nu(x)$ is the uncertainty of the GP model. The parameter $\beta$ controls the exploration-exploitation trade-off. Prioritizing a high $\mu$ and low $\nu$ by choosing a small $\beta$ in (\ref{eq:acq funct}), leads the algorithm to exploit known good results, while a high $\beta$ encourages exploration in areas with high uncertainty and potential reward. The full implementation of BO is summarized in Algorith \ref{alg:bayesian-optimization}.

\begin{algorithm}[H]
\caption{Bayesian Optimization Algorithm}
\label{alg:bayesian-optimization}
\begin{algorithmic}[1]
\Require Objective function $f(x)$, prior best data $\mathcal{G}_\text{max}$, acquisition function $\alpha(x)$, maximum iterations $N$
\Ensure Optimal parameters $x^*$ and $f^*$
\Statex \textbf{Note:} GP model uses a \texttt{ScaleKernel(Matern-5/2)} with ARD. Lengthscales are sampled from $\text{Gamma}(3.0, 6.0)$ and output scale from $\text{Gamma}(2.0, 0.15)$.
\State Initialize Gaussian Process (GP) $\mathcal{G} \gets \mathcal{G}_\text{max}$
\For{$n = 1, \ldots, N$}
    \State Compute posterior mean $\mu(x)$ and standard deviation $\nu(x)$ using GP
    \State Define acquisition function (UCB):
    \[
      \alpha(x) = \mu(x) + \beta \nu(x)
    \]
    \State Select next query point:
    \[
      x_n = \arg\max_{x} \alpha(x | \mathcal{G})
    \]
    \State Evaluate objective function: $y_n \gets f(x_n)$
    \State Update GP with new observation $(x_n, y_n)$:
    \[
        \mathcal{G} \gets \mathcal{G} \cup \{(x_n, y_n)\}
    \]
\EndFor
\State Find optimal parameters:
\[
    x^* = \arg\max_{x} f(x), \quad f^* = f(x^*)
\]
\State \Return $x^*, f^*$
\end{algorithmic}
\end{algorithm}

\section{\label{sec:results}Results}

Figure \ref{fig:plots} illustrates various aspects of optimizing the radioactive beamline Section 1. The top panel displays live optimization of the observed activity versus iteration. The times scale of our optimization iterations is selected such that, after background subtraction, an increment of the observed activity is proportional to an increment in the RIB intensity. Initially, the observed rate fluctuates as the algorithm explores the tuning space of quadrupoles and steerers, characterized by large changes. As iterations progress, the algorithm refines the machine configuration, resulting in steady RIB rate improvements. Despite significant drops around iterations 70, 80, and 100, the algorithm successfully returned to optimal settings. In the final iteration, a background measurement was conducted by removing the beam to measure the beam-off rate, stemming from the beta decay chain of the isotopes involved. This background is crucial for accurately determining the net activity of the beam. 

The Bateman equation~\cite{Bat1910} is a mathematical model used to describe the decay of isotopes in a sequential chain, where a parent isotope decays into a daughter isotope, which may further decay into subsequent isotopes. In our case, the equation is applied to a two-isotope chain, capturing the decay dynamics of the parent and daughter isotopes involved in the beta decay process.

The equation is given by:

\begin{figure}[b]
\centering
\includegraphics[width=0.8\linewidth]{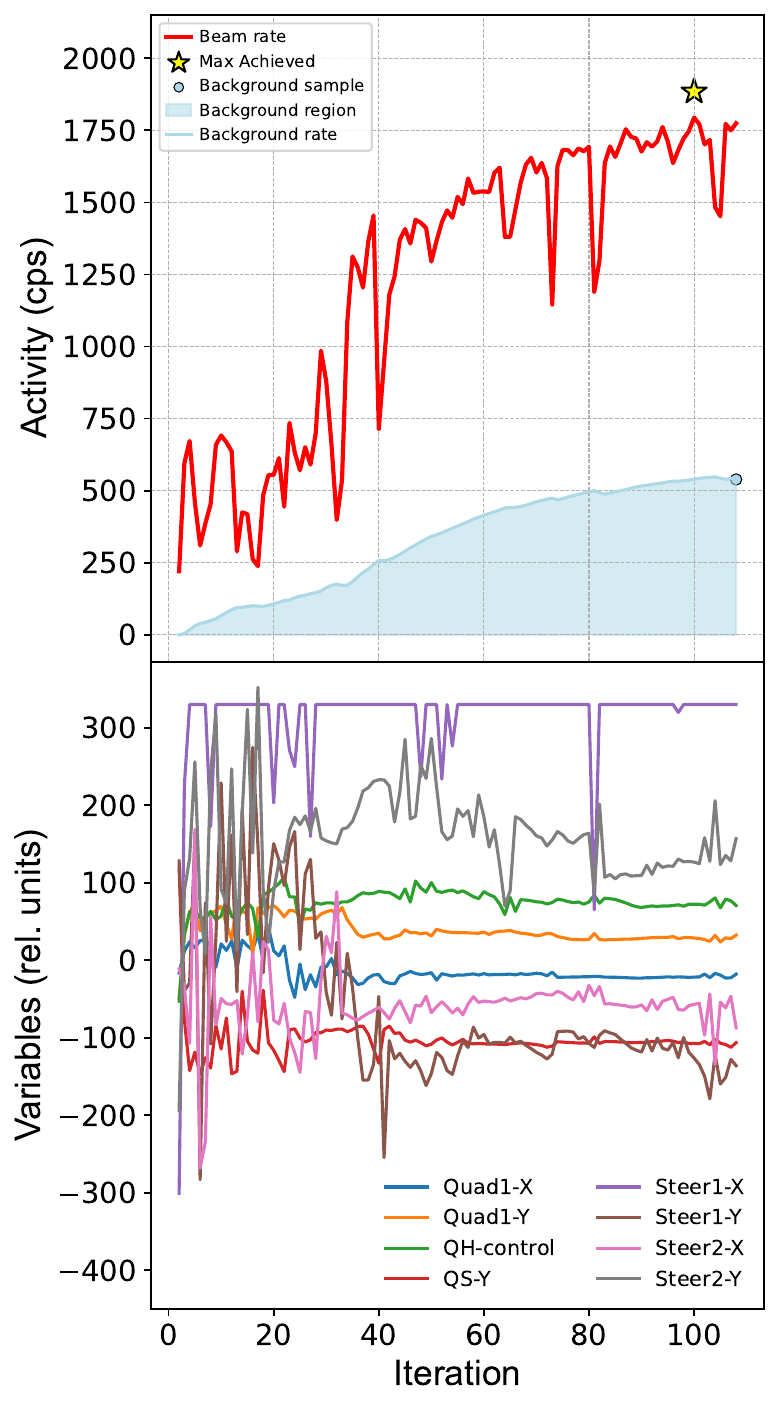}  
\caption{\justifying Live optimization of radioactive ion transport in beamline Section 1. In the top panel, the activity in counts per second (cps) is shown as a function of the optimization iteration. A yellow star marks the maximum, which corresponds to the optimum configuration. The background from beta decay, modeled using the Bateman equation, is shown in light blue. The bottom panel shows variable changes corresponding to the top panel activity. Each iteration took about 10 seconds.}
\label{fig:plots}
\end{figure} 

\begin{gather}
    A_B=\frac{\lambda_P}{\lambda_D-\lambda_P}A_P(0)(e^{-\lambda_Pt}-e^{-\lambda_Dt})+A_D(0)e^{\lambda_Dt}
\end{gather}

\noindent where $A_B$ is the background activity, $\lambda_P$ is the decay constant of the parent isotope, $\lambda_D$, the decay constant of the daughter isotope, $A_P$ is the initial beam rate, and $A_D$ is the initial daughter isotope rate. The Bateman equation allows us to predict the activity levels of these isotopes over time, providing a theoretical curve that represents the background activity due to beta decay. This curve, shown in light blue in the top panel of Fig. \ref{fig:plots}, is subtracted from the observed activity to calculate the net rate: (Observed Activity) - (Background Activity). By modeling the background activity accurately, we can isolate the true signal from the background. 

In the bottom panel of Fig.~\ref{fig:plots} the tuning of control variables across iterations highlights the balance between exploration and exploitation. Up to iteration 40, variations are large as the algorithm samples the parameter space. Later iterations converge towards optimal settings, demonstrating BO's capability in multi-dimensional optimization. 

Figure \ref{subfig:sensitivity} presents a correlation heatmap between tuning variables, providing insight into inter-variable influences and their relationship with the beam rate objective. To gain additional understanding of these relationships, we performed a post-hoc correlation analysis using data from the Bayesian Optimization process. Although the underlying Gaussian Process model employed ARD to learn per-variable sensitivities, our analysis computed Pearson correlation coefficients between each voltage setting and the resulting beam rate across all optimization steps. This correlation matrix, visualized as a heatmap, offers an intuitive overview of which variables exhibit the strongest linear associations with the beam rate.

Strong positive correlations (red) indicate direct proportionality, while negative correlations (blue) suggest inverse relationships. For instance, QH-control and Steer1-X have small positive correlations (0.336 and 0.280) with Beam Rate, indicating moderate increases with voltage. Conversely, Steer1-Y, Quad1-X, and Quad1-Y show high negative correlations (-0.706, -0.647, -0.547), suggesting beam rate benefits from reduced voltage. Elements like Steer2-X (-0.003), Steer2-Y (0.024), and QS-Y (-0.150) show near-zero correlation, indicating no direct influence on beam rate.

\begin{figure}[b]
\centering
\subfigure[\label{subfig:sensitivity}]{\includegraphics[width=0.89\linewidth]{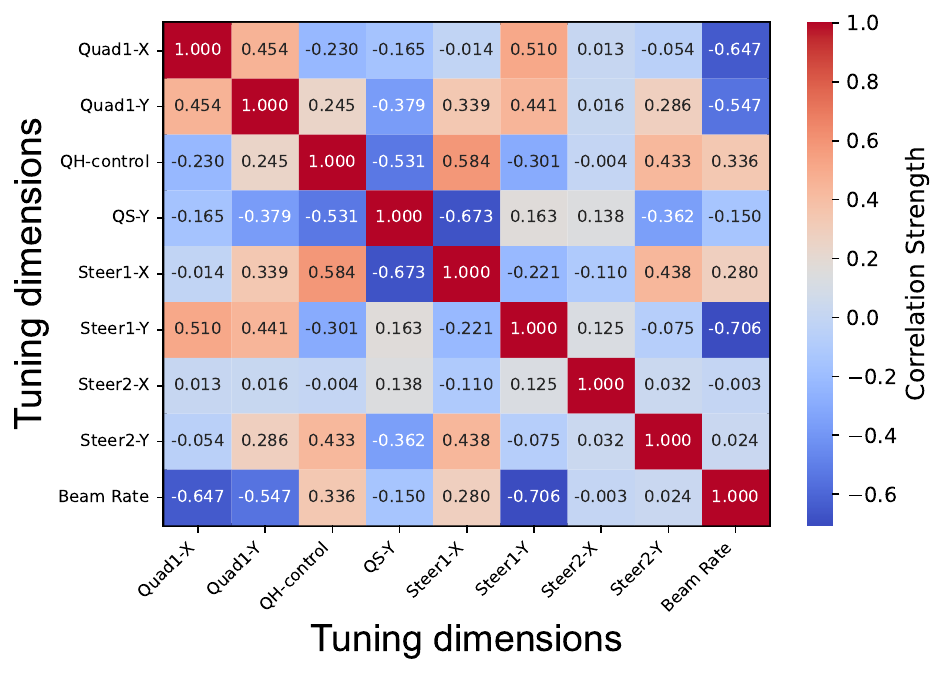}}

\subfigure[\label{subfig:3dplot}]{\includegraphics[width=0.75\linewidth]{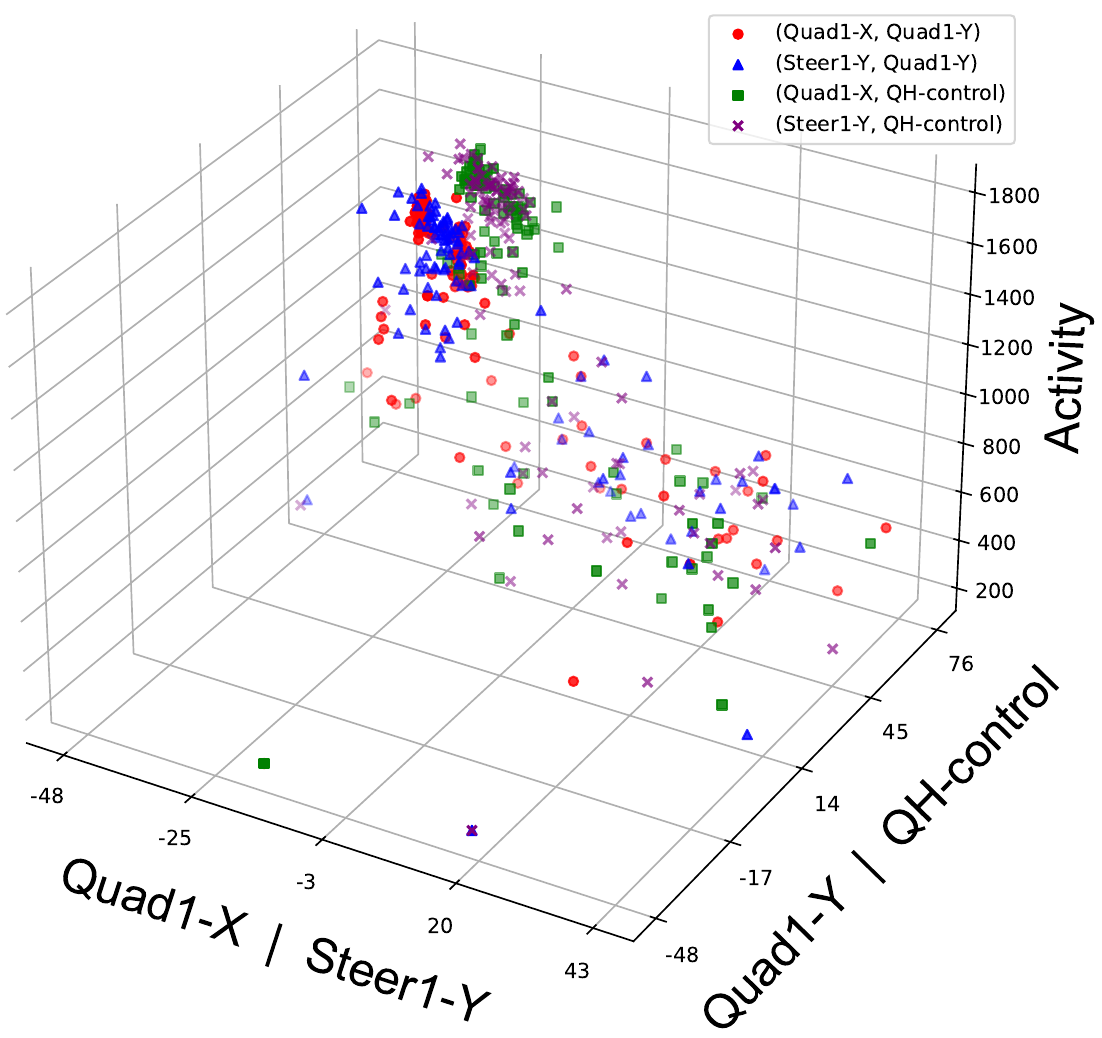}}

\caption{\justifying (a) Sensitivity heatmap showing the Pearson correlations between tuning parameters. This visualization highlights the linear associations between beam elements and beam rate. (b) 3D scatter plot illustrating trends among the most sensitive elements from (a). The best tuning region is identified by the highest concentration of points. Steer1-Y is normalized to Quad1-X, and QH-control is normalized to Quad1-Y for clearler visualization.}
\label{fig:plots-sensitivity}
\end{figure}

It is important to note that while ARD captures potentially nonlinear and interaction-based effects within the model, the correlation matrix reflects empirical, linear relationships within the observed data. Due to the non-uniform sampling nature of Bayesian Optimization—where the search is biased toward promising regions—the correlation values may not directly correspond to global sensitivities. However, strong correlations often indicate variables that contributed significantly to observed changes in beam rate, and can serve as a complementary tool to the ARD-learned lengthscales. 

While our correlation analysis provides empirical insights into linear relationships between tuning variables and beam rate, it is important to consider how this approach relates to existing methodologies. Hanuka et al. \cite{hanuka_physics_2021} explore the use of a physics-informed GP model for optimizing particle accelerators, embedding known physical correlations into the GP framework. This method enhances predictive capabilities by capturing interdependencies between input parameters, informed by physics-based simulations and modeling.

Our work aligns with these principles, utilizing empirical correlations as a complementary tool to ARD. By recognizing the role of correlated kernels, we highlight the potential for integrating physics-informed correlations into our analysis. This suggests avenues for future research to refine optimization strategies further, potentially combining empirical data-driven insights with physics-informed modeling to enhance the robustness of optimization algorithms.

Further analysis identifies Quad1-X (-0.647), Quad1-Y (-0.547), Steer1-Y (-0.706), and QH-control (0.336) as the most sensitive elements to beam rate. By focusing on these key elements, we can enhance optimization efficiency by reducing unnecessary variables, ultimately simplifying the troubleshooting process.

A 3D plot in Fig. \ref{subfig:3dplot} visualizes the aforementioned key variables versus beam rate, showing trends where high-intensity regions occur with specific parameter combinations. Optimal ranges include (-15 -- -25 V) for (Quad1-X, Steer1-Y) and (30 -- 45 V), (45 -- 65 V) for (Quad1-Y, QH-control). In this analysis, Steer1-Y is normalized to Quad1-X, and QH-control is normalized to Quad1-Y to bring these variables onto a comparable scale, facilitating clearer identification of relationships and trends. Concentration blobs in these regions result from proportionality differences. Simultaneous adjustments in the same direction for (Quad1-X, Quad1-Y) and (Steer1-Y, Quad1-Y) maintain favorable conditions, while inverse adjustments are needed for (Quad1-X, QH-control) and (Steer1-Y, QH-control). These results not only provide a comprehensive understanding of the optimization process but also pave the way for enhanced control strategies in RIB operations, potentially leading to more efficient and precise experimental setups. This advancement underscores the impactful potential of our approach in optimizing complex systems, including but not limited to RIB applications.

In our work, we explored the effect of varying the exploration-exploitation parameter $\beta$ in (\ref{eq:acq funct}) within the BO framework. This parameter plays a crucial role in balancing the trade-off between exploring new regions of the parameter space and exploiting known good configurations. Figure \ref{fig:beta} illustrates the impact of different $\beta$ values on the optimization process.

\begin{figure*}[t]
\centering  
\includegraphics[width=0.95\linewidth]{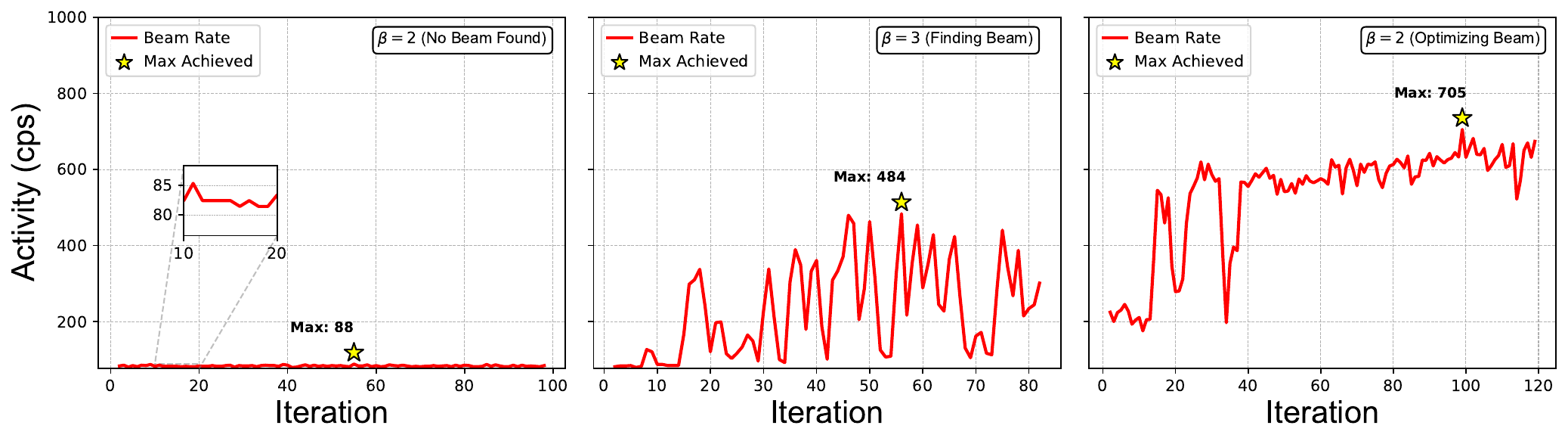}
\caption{\justifying A sequence of events showing how the RIB was found by varying the $\beta$ parameter.
Left: Initial case where no RIB is observed and a conservative exploration with $\beta=2$ stalls at a suboptimal local maximum. 
Center: Since no beam was found after about 100 iterations, a more aggressive exploration with $\beta=3$ was performed, restarting the optimization process from the same initial configuration as the previous run. 
In less than 20 iterations, the RIB was found, however relatively small gains were made in subsequent iterations. 
Right: After 80 iterations with $\beta=3$, the optimization was stopped and the parameter $\beta$ was returned to 2, while restarting the process with the variables from the last configuration where the RIB was observed. 
The optimization was resumed prioritizing exploitation, which resulted in a significant increment in the observed activity (i.e.~higher RIB intensity).}
\label{fig:beta}
\end{figure*}

Initially, a conservative approach with $\beta=2$ resulted in the algorithm stalling at a suboptimal local maximum, as no RIB was observed. By increasing $\beta$ to 3 and reinitializing the optimization process from a single starting configuration, the algorithm adopted a more aggressive exploration strategy. This change led the algorithm to successfully identify high-performance regions and detect the RIB. Subsequently, reverting to $\beta=2$ and reinitializing the process with the ``good-tune" parameters from the $\beta=3$ run, allowed the algorithm to focus on exploitation, converging to the global maximum and achieving an increased RIB rate.

The process described above for beamline Section 1 was applied in sequence through Section 5. While our AI-assisted RIB transport optimization achieved similar transmission efficiency ($\approx 50\%$) and transport time ($\approx$ 15 min per section) as manual methods, its true value lies in automating the tuning process. Unlike manual methods requiring constant parameter adjustments, our method automates beam delivery with minimal supervision, streamlining operations without sacrificing beam quality. As we expand the present method to other beamlines, it could significantly reduce tuning and setup time, allowing researchers to focus more on their scientific investigations.

\section{\label{sec:summary}Summary and Conclusion}
In this study, we have developed and tested an AI-assisted method for transporting radioactive ion beams at the CARIBU facility. The present methodology leverages Bayesian Optimization to automate the tuning process, typically reliant of manual adjustments by experienced scientists. This automation not only maintains the high transmission efficiency and transport times of manual methods but also significantly reduces the need for expert-level supervision. 

The successful implementation of our methodology at CARIBU demonstrates its potential for broader application across other beamline sections and similar facilities worldwide. By reducing setup time and optimizing beam tuning processes, our method empowers researchers to focus more on their scientific investigations rather than technical adjustments. This advancement marks a significant step towards autonomous scientific discovery in nuclear physics, enhancing workflow and empowering researchers to concentrate on addressing complex scientific questions.

Looking ahead, we plan an upgraded version of our approach for the Multi-Reflection Time-of-Flight (MR-TOF) device at CARIBU~\cite{Hir2016}. The MR-TOF device is a state-of-the-art mass spectrometer that exploits multiple reflections between electrostatic mirrors to extend the flight path of ions. This enables high-precision measurements of nuclear masses and efficient separation of isotopes during beam delivery operations. 

Beyond its success at specific facilities, BO's capabilities can be extended to domains facing similar high-dimensional tuning challenges, such as robotics, materials science, and hyperparameter tuning in machine learning models. In summary, our work not only advances the field of nuclear physics but also opens new avenues for applying AI-driven optimization in diverse scientific and technological fields. 

\begin{acknowledgments}
The authors wish to thank the ATLAS Controls Group for AWACS network access and CARIBU operators for facilitating live testing. This work was supported by the Department of Energy, Office of Science, Office of Nuclear Physics, under Contract No. DE-AC02-06CH11357, and DE-FOA-0002875 funds. This research used resources of ANL's ATLAS facility, which is a DOE Office of Science User Facility.
\end{acknowledgments}
\hspace{20em}

\bibliography{AI-assisted}

\end{document}